\documentclass[twoside,leqno]{article}
\usepackage[letterpaper]{geometry}
\usepackage{siamproceedings}
\usepackage[utf8]{inputenc}
\usepackage[T1]{fontenc}

\hypersetup{
    hidelinks,
    pdftitle={Nearly Tight Lower Bound for Range Minimum Query via Probe-by-Probe Information Accounting},
    pdfauthor={Mingmou Liu}
}
\usepackage[algo2e,boxruled]{algorithm2e}
\numberwithin{algocf}{section}
\crefname{algocf}{Algorithm}{Algorithms}
\Crefname{algocf}{Algorithm}{Algorithms}

\usepackage{amsfonts}
\usepackage{amssymb}
\usepackage{tikz}
\usepackage{mathrsfs}
\usepackage[colorinlistoftodos]{todonotes}
\usepackage[inline]{enumitem}
\usepackage{booktabs}
\usepackage{tabularx}

\usepackage{thmtools}
\usepackage{thm-restate}
\newsiamthm{claim}{Claim}
\newsiamthm{fact}{Fact}
\newsiamthm{protocol}{Protocol}
\AddToHook{env/protocol/begin}{\crefalias{theorem}{protocol}}
\crefname{claim}{Claim}{Claims}
\Crefname{claim}{Claim}{Claims}
\crefname{fact}{Fact}{Facts}
\Crefname{fact}{Fact}{Facts}

\newcommand{\E}{\mathop{\mathbb{E}}}

\newcommand{\HammingCube}[1]{\{0,1\}^{#1}}

\DeclareMathOperator*{\argmin}{arg\,min}

\newcommand{\pub}{\mathrm{pub}}
\newcommand{\probe}{\mathtt{Probe}}
\newcommand{\pred}{\mathtt{pred}}

\newcommand{\rmq}{\texttt{RMQ}}
\newcommand{\cQ}{\mathcal{Q}}
\newcommand{\cD}{\mathcal{D}}
\newcommand{\predz}{\texttt{pred-z}}
\newcommand{\pt}{\mathrm{pt}}
\newcommand{\cS}{\mathcal{S}}
\newcommand{\cE}{\mathcal{E}}
\newcommand{\cK}{\mathcal{K}}
\newcommand{\cG}{\mathcal{G}}
\newcommand{\rest}{\mathtt{remain}}
\newcommand{\good}{\mathrm{good}}
\newcommand{\foot}{\mathtt{Foot}}

\begin{document}
\title{\Large Nearly Tight Lower Bound for Range Minimum Query \\ via Probe-by-Probe Information Accounting}
\author{Mingmou Liu\thanks{School of Intelligent Software and Engineering,
    Nanjing University, China (\email{lmm@nju.edu.cn}).}}
\date{}
\maketitle

\fancyfoot[C]{\thepage}

\begin{abstract}
We prove a nearly tight lower bound for succinct Range Minimum Query (RMQ) on an array of $n$ distinct integers.
A query returns the position of the smallest entry in a specified subarray.
For word size $\Theta(\log n)$, any cell-probe data structure answering queries in time $t$ requires
$2n-1.5\log n+n/(\log n)^{O(t\log^2 t)}$ bits of space.
This improves the exponent $O(t^2\log^2 t)$ in the lower bound of Liu and Yu (STOC 2020), leaving only a polylogarithmic gap in query time from the known upper bound across the redundancy range.
The key ingredient is \emph{probe-by-probe information accounting}.
The previous analysis measures useful information at an ``average'' memory-access round of an adaptive query algorithm.
For RMQ, the informative queries are database-dependent and unevenly distributed, making the conversion from this information to query savings lose a factor of $t$.
We account for information across all memory accesses using an entropy telescoping argument.
This gives a stronger progress bound within the existing published-bit round-elimination framework.
The resulting viewpoint suggests a way to address similar averaging losses in other succinct lower bounds, although applications beyond RMQ remain open.
\end{abstract}

\section{Introduction.}
How many memory words must a data structure read to answer a query?
In the cell-probe model~\cite{yao1981should}, computation is free and only memory accesses, traditionally called \emph{cell probes}, are counted.
Because computation is unrestricted, a lower bound must hold even for algorithms with unlimited computational power between memory accesses.
A cell-probe lower bound is \emph{unconditional}: it requires no computational hardness conjecture and also lower bounds word-RAM query time under the space and word-size constraints.

A \emph{succinct} data structure supports queries using space close to the information-theoretic minimum.
Such representations help maintain large datasets with small space and are a basic tool for compact text indexes and trees~\cite{Sada07b,NS14}.
Practical examples of querying compressed representations include the Succinct data store~\cite{AgarwalKhandelwalStoica15} and the Elias--Fano encoded lists in Meta's Folly library~\cite{MetaFollyEliasFano}.
They also separate the cost of storing information from the cost of accessing it: enough bits to determine every answer need not suffice for fast queries.
The central question is how much extra space above the information-theoretic minimum, or \emph{redundancy}, is necessary to achieve a given query time.

The published-bit round-elimination method of P\v{a}tra\c{s}cu and Viola~\cite{PV10} is a principal tool for proving such trade-offs for static succinct data structures.
It recursively makes a small amount of memory available for free and argues that queries can then skip some memory accesses.
Liu and Yu~\cite{liu2020lower} adapted this method to Range Minimum Query (RMQ), obtaining a nearly tight lower bound for constant query time.
For super-constant query time, however, a substantial gap remained.
We close this gap to within polylogarithmic factors by introducing a finer analysis of how adaptive queries acquire information.

\paragraph*{Range Minimum Query.}
Given an array of distinct integers $A[1\ldots n]$, a query specifies an interval $[a,b]\subseteq[1,n]$ and returns
\[
    \rmq(a,b)\triangleq\argmin_{i\in[a,b]} A[i].
\]
RMQ is a basic primitive in graph algorithms~\cite{RV88,GT04}, text processing~\cite{ALV92,FHK06,Sada07}, and other algorithmic applications.
It also provides a natural test case for succinct lower bounds: the Cartesian tree~\cite{10.1145/358841.358852} of $A$ captures exactly all its RMQ answers.
There are $C_n=\binom{2n}{n}/(n+1)$ such trees, so the information-theoretic space requirement is $\log C_n=2n-1.5\log n+O(1)$ bits.

\newpage
\subsection{Result and comparison with prior bounds.}
Following a line of compact and succinct representations~\cite{GBT84,HT84,Sada07b,FH07,FH11}, the augmented-B-tree construction of Navarro and Sadakane~\cite{Pat08,SN10,NS14} gives the upper bound in \Cref{tab: rmq bounds}.
The table compares redundancy above $2n-1.5\log n$ at query time $t$, up to constant factors in time, for word size $\Theta(\log n)$.
Constant additive terms in the entropy baseline are suppressed.
As usual for this upper bound, a precomputed lookup table independent of the input is treated separately.
\begin{table}[ht]
\centering
\begin{tabularx}{\linewidth}{@{}>{\centering\arraybackslash}X >{\centering\arraybackslash}X >{\centering\arraybackslash}X@{}}
\toprule
Source & Type & Redundancy $r$ \\
\midrule
Navarro--Sadakane~\cite{NS14} & Upper bound &
$\displaystyle n\bigg/\left(\frac{\log n}{t}\right)^t$ \\[6pt]
Liu--Yu~\cite{liu2020lower} & Lower bound &
$\displaystyle n/(\log n)^{O(t^2\log^2 t)}$ \\[6pt]
This paper & Lower bound &
$\displaystyle n/(\log n)^{O(t\log^2 t)}$ \\
\bottomrule
\end{tabularx}
\caption{Space--time trade-offs for succinct RMQ. The new bound removes a factor of $t$ from the previous lower-bound exponent.}
\label{tab: rmq bounds}
\end{table}

Our lower bound holds generally for word size $w=\Omega(\log n)$.
\begin{restatable}{theorem}{thmmainconts}
    \label{thm_main}
    In the cell-probe model with word size $w=\Omega(\log n)$, any data structure supporting \rmq{} on an array of $n$ distinct integers with $2n-1.5\log n+r$ bits of space and query time $t$ must satisfy
    \[
        r\ge n/w^{O(t\log^2 t)}.
    \]
\end{restatable}
To compare the bounds at the same redundancy, fix $w=\Theta(\log n)$ and let $t_{\mathrm{NS}}(r)$ denote the query-time guarantee of the Navarro--Sadakane upper bound~\cite{NS14} at redundancy $r$, under the lookup-table convention above.
Our theorem gives
\[
    t=\Omega\!\left(\frac{t_{\mathrm{NS}}(r)}
    {\log^3(2+t_{\mathrm{NS}}(r))}\right).
\]
Thus the gap between the stated upper and lower bounds is at most a cubic logarithmic factor in the upper-bound query time.
For example, with $O(1)$ bits of redundancy, our time lower bound is
$\Omega(\log n/(\log\log n)^3)$, compared with the known $O(\log n)$ time upper bound.
The result is nearly tight; a polylogarithmic gap remains.

\paragraph*{The new ingredient.}
Our key ingredient is \emph{probe-by-probe information accounting}.
It tracks the information acquired throughout a query algorithm's adaptive sequence of memory accesses, rather than estimating progress from a single average memory-access round.
Combining these contributions gives stronger query-time savings in each round-elimination step, yielding the nearly tight RMQ bound.
\Cref{section: technique} explains the obstruction and the new argument without assuming familiarity with Liu and Yu~\cite{liu2020lower}.

\subsection{Related work.}
The closest line of work concerns static succinct lower bounds.
P\v{a}tra\c{s}cu and Viola~\cite{PV10} introduced the published-bit method for rank; Yu~\cite{Yu19} gave an optimal succinct rank construction.
Liu and Yu~\cite{liu2020lower} developed the distribution and reduction used here; succinct accessible interval graphs provide another application of round elimination~\cite{chakraborty2023cellprobe}.
Other approaches include Golynski's verification-based method~\cite{Gol09} and Viola's separator method~\cite{viola2021lower}.
The latter handles problems such as rank and colored multi-predecessor, whose datapoints are evenly distributed; the uneven distribution arising from RMQ is the obstacle relevant to this paper.

Related time-space trade-offs arise in hash tables~\cite{BenderFarachColtonKuszmaulKuszmaulLiu22}, dynamic succinct dictionaries~\cite{LiLiangYuZhou23}, and monotone minimal perfect hashing~\cite{AssadiFarachColtonKuszmaul25,Kosolobov24}.
Dynamic lower bounds, including those of Larsen, Weinstein, and Yu~\cite{LWY20} and Ko~\cite{Ko25,Ko26}, address a different regime.
Our focus here is the interaction between adaptivity and unevenly distributed queries in the succinct RMQ argument.

\section{Technique Overview: Probe-by-Probe Information.}\label{section: technique}
We first introduce the model and the predecessor problem used by the proof.
We then explain the compression argument behind published-bit round elimination, the difficulty caused by adaptive memory accesses and unevenly distributed hard queries (which are exactly the datapoints in case of RMQ), and how the new accounting avoids the resulting loss.

\subsection{The model and the predecessor problem.}\label{section: model}
Throughout, $[n]\triangleq \{1,\ldots,n\}$ and information quantities use base-two logarithms.
We write $H(X)$ for Shannon entropy and $I(X:Y)$ for mutual information.
Logarithms in asymptotic bounds are truncated below at $1$.

\paragraph*{Memory and published bits.}
A cell-probe data structure stores $s$ memory cells of $w$ bits each.
A query algorithm adaptively reads at most $t$ memory cells: the address of each access can arbitrarily depend on the query and on the contents already read.
A memory access is traditionally called a \emph{cell probe}, hence the name of the model.
All other computation is free.
The published-bit variant additionally allows $p$ published bits, depending on the database, to be read for free.
For a database universe $\mathcal B$, the representation is a map
$ T:\mathcal B\longrightarrow(\HammingCube{w})^s\times\HammingCube{p}.  $
If a representation uses $\log_2|\mathcal B|+r$ bits, its redundant bits can be placed in the published part.
The lower-bound argument therefore works with a memory of information-theoretically optimal size and a small published part.
A \emph{round-elimination} step publishes more information to reduce query cost.
Within such a step, an \emph{access round} refers to the $i$-th memory access of each query, not to another step of the recursion.

\paragraph*{The problem \predz{}.}\label{sec: predz definition}
For a set $S$ of integers, the predecessor of $x$ is the largest element of $S$ not exceeding $x$.
The problem \predz{} is a variant of predecessor search over multiple sets, with an input consisting of $d$ sets
$S_1,\ldots,S_d\subseteq[B]$, each with $u$ elements, together with an auxiliary integer $z$.
A query $\pred(i,x)$ asks for the predecessor of $x$ in the specified set $S_i$, returning
$ \max\bigl(\{0\}\cup(S_i\cap[1,x])\bigr).$
The structure must also support \texttt{query-z}(), which returns $z$.
We bound the time for predecessor queries only; \texttt{query-z} may take arbitrarily long.
$z$ and \texttt{query-z}() together serve to reduce \predz{} to \rmq{}.
Specifically, \texttt{query-z} requires the data structure to store extra information so that the entropy gap between \predz{} and \rmq{} is small enough.
The allowed range of $z$ depends on the sets and a parameter $Z$, defined in \Cref{sec: exact distribution}.

\paragraph*{The inherited hard distribution and reduction.}\label{sec: hard distribution}
Liu and Yu~\cite{liu2020lower} construct a distribution $\cD$ in which the $d$ predecessor sets are mutually independent.
The weights of $\cD$ on inputs reflect the tree structures counted by a uniformly random Cartesian tree, rather than uniformly random datapoints.
Intuitively, the predecessor sets describe tree paths, while $z$ encodes the remaining structure.
With the parameters used in the reduction, the full input entropy is $2n-\Theta(d\log B)\approx 2n-\Theta(r\log n)$, where $r$ is the redundancy of a succinct data structure we are proving lower bound for, close to the RMQ entropy $\log_2 C_n=2n-1.5\log n+O(1)$.
An RMQ structure with $r$ bits of redundancy can answer the corresponding predecessor queries, with redundancy $O(d\log B)\approx O(r\log n)$ bits above the \predz{} entropy.
It therefore suffices to prove a lower bound for \predz{} under $\cD$.
The formal distribution and the reduction are recalled in \Cref{sec: exact distribution,sec: reduction}.

An important feature of $\cD$ is its uneven spacing of datapoints: large empty gaps coexist with dense clusters.
Queries spaced uniformly across the entire universe need not expose the information needed for a strong lower bound.
This unevenness, together with adaptivity, is the obstacle addressed below.

\subsection{The inherited round-elimination framework.}\label{section: ly technique}
The recursive move in the method of P\v{a}tra\c{s}cu and Viola~\cite{PV10} is to choose a small \emph{public query set} $Q_{\pub}$ and publish the addresses and contents of all memory cells accessed when answering those queries.
Write $\probe(Q)$ for the set of addresses of the memory cells accessed when answering a query family $Q$.
We call the memory cells at addresses in $\probe(Q_{\pub})$ the \emph{public memory cells}.
If another query accesses one of these memory cells, it can now obtain the contents from the published bits and skip that memory access.
The goal is to prove substantial overlap between $\probe(q)$ and $\probe(Q_{\pub})$ for an average query $q$, so that each publishing step reduces query cost, i.e. saves memory accesses.
Repeating this step yields a contradiction if the original redundancy is too small: at the end of the recursion, there is not memory-accessing, and all queries can be answered with only the published bits, but the number of published bits is smaller than the entropy of the database.

For rank~\cite{PV10}, which asks for the number of $1$-bits in a prefix of a bit vector, evenly spaced public queries expose prefix sums that share information with answers to typical queries.
These query locations can be chosen independently of the database.
If the two query families accessed too few common memory cells, their shared answer information would be represented twice in nearly disjoint parts of memory.
A compression argument turns this duplication into savings exceeding the redundancy, producing an encoding shorter than the input entropy.
This forces the overlap needed for round elimination.

\newpage
\paragraph*{Database-dependent informative queries.}
For RMQ, many intervals have the same minimum; for example, every interval containing the global minimum returns its position.
A database-independent random query can therefore be much less informative than a database-dependent query.
Therefore the public queries $Q_{\pub}$, and the carefully chosen hard queries $\cQ$ to prove time cost lower bounds, both must be database-dependent.
Writing down the informative queries during a compression argument could cost more than the compression could save.
Moreover, even after a public query set has been specified, the addresses accessed when answering it depend on the database through the query algorithm's adaptive memory accesses.
Thus neither the hard queries nor the accessed memory addresses can simply be treated as database-independent.

Liu and Yu~\cite{liu2020lower} choose an informative public query set without describing every hard query.
Given a succinct data structure with $p$ published bits, Liu and Yu~\cite{liu2020lower} specify public queries costs $O(p\log B)$ bits, whereas their answers reveal $\Omega(p\log^2 B)$ bits of information.
The compression argument can afford this description cost, provided it subsequently saves $\omega(p\log B)$ bits.

Two consecutive public queries at positions $a<b$ reveal whether $(a,b]$ contains a datapoint:
\[
    S_i\cap(a,b]\ne\emptyset
    \quad\Longleftrightarrow\quad
    \pred(i,a)\ne\pred(i,b).
\]
Let $\cE$ be the set of these non-empty intervals.
Thus the answers to $Q_{\pub}$ determine $\cE$.
Liu and Yu show that this occupancy pattern has high entropy.
The formal construction and the precise inherited estimates appear in \Cref{sec_qpub,sec: inherited estimates}.

\paragraph*{Shared information does not immediately force overlap.}
Suppose most hard queries access no memory cell at an address in $\probe(Q_{\pub})$.
Their answers can then be recovered using only the contents of memory cells at addresses outside $\probe(Q_{\pub})$.
The decoder need not know the hard queries: it can try all predecessor queries and stop any execution that attempts to access a public memory cell.
The intended compression uses the information about $\cE$ shared by these recoverable answers and the public answers.
If the memory partition were fixed independently of the database, sufficiently large shared information would allow one part of memory to compress the other, giving an encoding shorter than the database entropy.
The decoder, however, must know which addresses are outside $\probe(Q_{\pub})$ before it can run those simulations.
For an adaptive query algorithm, membership in $\probe(Q_{\pub})$ can itself carry information about the input.
Explicitly describing all these addresses could consume the intended compression savings.
The proof must therefore track the information in the query algorithm's adaptive memory-access sequence, not merely the overlap of two sets of addresses.

\subsection{The obstacle: single-round averaging under uneven distribution.}
Let the \emph{public footprint} $F$ denote the sequence of memory-cell contents read while answering $Q_{\pub}$.
Once the public queries and published bits are given, these contents determine the adaptive memory-access sequence used to answer each public query, and hence the address set $\probe(Q_{\pub})$ and the answers to $Q_{\pub}$.
In particular, it determines the non-emptiness information in $\cE$.
Liu and Yu~\cite{liu2020lower} analyze this information by isolating one memory-access round $l$ and conditioning on the memory contents read in earlier rounds.
The prefix (i.e. the memory contents read in earlier rounds) determines the addresses accessed in round $l$, so any additional shared information in round $l$ is information revealed by memory-cell contents alone, usable for compression without any extra address description.
The chain rule distributes the useful information over all the $t$ rounds; a uniformly selected round accounts for a $1/t$ share in expectation.

To turn this information estimate into query-time savings, Liu and Yu~\cite{liu2020lower} use interval non-emptiness as a witness.
A hard query at a datapoint in an interval returns that datapoint, certifying that the interval is non-empty.
The argument must connect this witness information to overlap between the memory cells accessed by hard and public queries, as both answers to them reveal same emptiness information $\cE$.
Under $\cD$, interval occupancies are highly uneven: a non-empty interval can contain very few or very many datapoints, each corresponding to a hard query.
The information assigned to the selected memory-access round may concern sparse intervals, while other rounds concern intervals containing many more hard queries.
The conversion is in fact showing that every datapoint, i.e. every hard query, inside the intervals associated with round $l$ has to access a memory cell in $\probe(Q_{\pub})$.
Therefore the unevenness might result in only a small query-time saving per elimination step.
In Liu and Yu's analysis, the resulting guarantee is only $\Omega(1/(t\log^2 t))$ expected memory accesses saved per elimination step.
Repeated over the query's lifetime, this leads to the exponent $O(t^2\log^2 t)$.
The obstruction is this combination of adaptivity, single-round averaging, and uneven query density.

\subsection{Accounting across all memory accesses.}\label{sec: overview accounting}
Our approach is to account for what each successive memory access reveals and combine these contributions over the query algorithm's entire adaptive memory-access sequence.
We adopt Liu and Yu's hard distribution $\cD$ and query families $\cQ$ and $Q_{\pub}$, but change how their information is analyzed.
We estimate the total interval information carried by all memory-access rounds before converting it into overlap between the sets of memory-cell addresses accessed by public and hard queries.
Thus we need not assume that any particular memory-access round carries a typical share of the useful information or that its intervals have typical occupancies.
For RMQ, this avoids the weak conversion forced by a single selected round.

\paragraph*{Counting hard queries via unknown intervals.}
Let $\cK'$ be the intervals in $\cE$ containing a datapoint whose corresponding hard query \emph{does not} access any public memory cell.
Then every hard query at a datapoint in $\cE\setminus\cK'$ accesses a public memory cell.
A lower bound on the number of these intervals, combined with their datapoint counts, therefore gives a lower bound on the number of hard queries that benefit from publishing.
This then gives the expected number of memory accesses saved per elimination step.

Estimating $|\cE\setminus\cK'|$ directly is hard, so we use a larger certifiable set $\cK\supseteq\cK'$.
An interval belongs to $\cK$ if some predecessor query returns a datapoint in it without accessing any public memory cell; the certifying query need not be hard.
We prove a lower bound on $|\cE\setminus\cK|\le |\cE\setminus\cK'|$ instead.
The set $\cE\setminus\cK$ consists of non-empty intervals whose non-emptiness cannot be certified by a predecessor query that avoids the public memory cells.
Following Liu and Yu~\cite{liu2020lower}, we call these \emph{unknown intervals}.
Intuitively, they represent interval information revealed through the public footprint $F$.
Our main lemma proves
\[
    \mathbb E[|\cE\setminus\cK|]=\Omega\left(\frac{p\log^2 B}{\log t}\right).
\]
The inherited interval-to-query conversion then saves $\Omega(1/\log^2 t)$ expected memory accesses per elimination step, improving Liu and Yu's guarantee by a factor of $t$.

In the entropy expressions below, we condition on the published bits and the data specifying $Q_{\pub}$, and suppress this common side information from the notation.
We estimate the size of $\cE\setminus\cK$ through the conditional entropy $H(\cE\mid\cK)$. %
Since $H(\cE)=I(\cE:\cK)+H(\cE\mid\cK)$ and $H(\cE)=\Theta(p\log^2B)$, it remains to control the mutual-information term in terms of the same count $|\cE\setminus\cK|$.
On the other hand, note that $\cE$ is determined by $F$ and $Q_{\pub}$ due to the adaptive query algorithm, so we think about controlling $I(F:\cK)$ instead.

\paragraph*{Accounting one footprint layer at a time.}
Given the public queries and the published bits, the contents $F$ determine all their answers and hence $\cE$.
Let $F_i$ denote the contents read at the $i$-th memory access of each public query.
The prefix footprints $F_{<i}$ and $F_{\le i}$ are the contents of the memory cells accessed by public queries before and through round $i$, respectively.

The probe-by-probe information accounting starts from the chain rule
\begin{align}
	H(F_{\le t}|\cK)=H(F_t\mid F_{\le t-1},\cK)+H(F_{\le t-1}\mid \cK),\label{eq: chain rule}
\end{align}
and peels off one layer $F_i$ at a time while moving the certifiable set $\cK$ in sync.
For $i=0,\ldots,t$, let $\cK_i$ be the non-empty intervals that remain certifiable when the memory cells accessed in the first $i$ rounds of $Q_{\pub}$ are unavailable, so
\[
	\cK=\cK_t\subseteq\cK_{t-1}\subseteq\cdots\subseteq\cK_0=\cE .
\]
The quantity $H(F_{\le i}\mid \cK_i)$ 
allows us to apply the chain rule in \cref{eq: chain rule} recursively: the footprint prefix shrinks from $F_{\le i}$ to $F_{\le i-1}$, while the certifiable set expands from $\cK_i$ to $\cK_{i-1}$.
One local estimate says that the conditional entropy $H(F_i\mid F_{\le i-1},\cK_i)$ is almost the ordinary increment $H(F_i\mid F_{\le i-1})$.
The other local estimate replaces the condition $\cK_i$ by $\cK_{i-1}$, paying only for the local change from $\cK_{i-1}$ to $\cK_i$.
After summing over all $i$, these inequalities telescope into
\[
    H(F\mid \cK)\ge H(F)-\sum_{i=1}^t H(\cK_i,\cK_{i-1}\setminus\cK_i\mid \cK_{i-1}) -O(tp\log B).
\]
Equivalently,
\[
	I(F:\cK)\le
	\sum_{i=1}^t H(\cK_i,\cK_{i-1}\setminus\cK_i\mid \cK_{i-1}) +O(tp\log B).
\]
Since $\cE$ is determined by $F$, the data processing inequality gives $I(\cE:\cK)\le I(F:\cK)$.
Together with the identity $I(\cE:\cK)=H(\cE)-H(\cE\mid\cK)$, this yields
\[
	    H(\cE) \le H(\cE\mid\cK)+\sum_{i=1}^t H(\cK_i,\cK_{i-1}\setminus\cK_i\mid \cK_{i-1}) +O(tp\log B).
\]
The large entropy of the non-empty intervals $H(\cE)=\Omega(p\log^2B)$ must be accounted for either by the unknown part $H(\cE\mid\cK)$ or by the accumulated local changes along the chain $\cK_0,\cK_1,\ldots,\cK_t$, apart from the last term.
Both the first two terms can be bounded in terms of $\mathbb E[|\cE\setminus\cK|]$ from above.
This forces $\cE\setminus\cK$ to be large, which in turn forces many hard queries to access the public memory cells.

\subsection{Scope of the idea.}
A natural approach to lower bounds for succinct data structures is to identify informative queries and the corresponding footprint.
A query algorithm with adaptive memory accesses accumulates observations; our proof tracks the adaptive probe-by-probe information accumulation using the complementary perspective of progressively making the memory cells accessed in successive rounds unavailable and tracking the certificates that survive.
This suggests a way to handle other problems with database-dependent, unevenly distributed hard queries.
The appropriate answer information, certificates, and quantitative entropy and encoding estimates still have to be established separately for each problem.
We prove the RMQ application here; further applications remain open.

\subsection{Proof organization and provenance.}
The reduction, hard distribution, public-query construction, and distributional estimates come from Liu and Yu~\cite{liu2020lower}.
The new part is the joint analysis of the footprint and the certificate chain, proved in \Cref{sec_one_round}.
The imported estimates are collected in \Cref{sec: inherited estimates}.
\Cref{sec_lower} converts the new bound into memory-access savings and applies the inherited recursion and reduction.
The encoding-length proof, which follows Liu and Yu closely, is included in \Cref{section: lem_enc_lem}.

\section{The New One-Round Lemma via Probe-by-Probe Information Accounting.}\label{sec_one_round}
The new task is to upper bound $I(\cE:\cK)$, the information shared between the intervals exposed by the public queries and those still certifiable without accessing the public memory cells, in terms of $\mathbb E[|\cE\setminus\cK|]$.
The reason is the identity
\[
    H(\cE)=I(\cE:\cK)+H(\cE\mid\cK).
\]
The inherited estimates make $H(\cE)$ large and bound $H(\cE\mid\cK)$ in terms of $|\cE\setminus\cK|$.
Our accounting bounds the mutual-information term in terms of the same count.
If $\cE\setminus\cK$ is too small, the two terms could not account for $H(\cE)$, resulting in a contradiction.

We use the problem and distribution $\cD$ described in \Cref{section: model}, with $u=\Theta(\sqrt B)$.
Let $H_{d,u,B,Z}$ denote the entropy of the full \predz{} input under $\cD$, equivalently the logarithm of the number of valid inputs; its exact formula is given in \Cref{sec: exact distribution}.
All data structures in this section have at most $H_{d,u,B,Z}$ bits of memory in addition to their published bits.
The symbols $Q_{\pub}$, $\cE$, and $\cK$ have the meanings explained in \Cref{section: technique}; their exact construction is given in \Cref{sec_qpub}.
\begin{restatable}{lemma}{lemunknowninterval}
    \label{lemma: unknown intervals}
    Suppose a \predz{} data structure has at most $H_{d,u,B,Z}$ bits of memory, $p$ published bits with $d\le p<du/\log^4 B$, and worst-case predecessor query time $t=o(\log B)$.
    Then there is a public set $Q_{\pub}$ of $O(p\log^4 B)$ queries such that
    \[
        \mathbb E[|\cE\setminus\cK|]=\Omega(p\log^2 B/\log t).
    \]
\end{restatable}
The expectation is taken over the input drawn from $\cD$ and a random shift used to choose the public queries.
We first explain why the proof uses the footprint and the nested certificate sets, then give the formal construction, collect the estimates inherited from Liu and Yu~\cite{liu2020lower}, and prove the new accounting.

\subsection{Intuition for the accounting.}\label{sec: accounting intuition}
The set $\cE$ consists of the non-empty intervals between consecutive public queries, while $\cK$ consists of those that are still certifiable without accessing the public memory cells.
Intuitively, $\cE\setminus\cK$ is the interval information that must have passed through the memory cells accessed by $Q_{\pub}$.

Technically, we would like to lower bound $|\cE\setminus\cK'|$, where $\cK'\subseteq\cK$ consists of intervals containing a datapoint whose corresponding hard query avoids all public memory cells.
Every hard query at a datapoint in $\cE\setminus\cK'$ therefore accesses a memory cell at an address in $\probe(Q_{\pub})$.
Counting these intervals and their datapoints gives a lower bound on the number of hard queries that benefit from publishing, and hence on the expected memory accesses saved per elimination step.
Since estimating $|\cE\setminus\cK'|$ directly is hard, we lower bound the smaller quantity $|\cE\setminus\cK|$ instead.

Directly examining $\cE\setminus\cK$ is hard.
A useful detour is that all information in $\cE$ is contained in the public footprint $F$: once the answers to $Q_{\pub}$ are known, $\cE$ is known, and these answers are functions of $F$.
We therefore measure the interaction between $F$ and $\cK$ first, and only later translate it back to $\cE$.

Recall that $F_i$ is the $i$-th layer of the public footprint and $F_{<i}$ is the prefix before it.
Start with $H(F\mid\cK)$.
The chain rule decomposes it from the last access round as
\begin{align}
	H(F\mid\cK)=H(F_t\mid F_{<t},\cK)+H(F_{<t}\mid\cK).
	\label{eq: chain rule2}
\end{align}
The intended accounting is the following.
The first term should be compared with the incremental entropy $H(F_t\mid F_{<t})$, without the additional condition $\cK$.
If this comparison can be repeated for every round $i\le t$, then the increments $H(F_i\mid F_{<i})$ over $i\in[t]$ add up to $H(F)$, a quantity we can handle.
We also treat the second term recursively, following the same pattern as Eq\cref{eq: chain rule2}.
Here $\cK$ denotes the set of non-empty intervals that remain certifiable when all public memory cells are unavailable.
The conditioning set $\cK$ should move in sync with the footprint prefixes $F_{\le t},F_{\le t-1},\ldots,F_{\le0}$.
This is why we introduce the nested sets
\[
	\cK=\cK_t\subseteq\cK_{t-1}\subseteq\cdots\subseteq\cK_1\subseteq\cK_0=\cE,
\]
where $\cK_i$ is the interval information still certifiable when the memory cells accessed in the first $i$ memory-access rounds of $Q_{\pub}$ are made unavailable.
While $F_{\le i}$ shrinks to $F_{<i}$ in the entropy calculation, the side information should expand from $\cK_i$ to $\cK_{i-1}$ simultaneously.

The two claims in \Cref{sec: local entropy} implement these two comparisons.
\Cref{claim: tail extent} shows that the first term is almost the ordinary increment $H(F_i\mid F_{<i})$.
\Cref{claim: condition manipulate} changes the second term from conditioning on $\cK_i$ to conditioning on $\cK_{i-1}$, paying only the entropy needed to describe the local change from $\cK_{i-1}$ to $\cK_i$.
Write $\cK'_{i-1}=\cK_{i-1}\setminus\cK_i$ for the non-empty intervals that disappear at this step.
Thus the proof accumulates the quantities
\begin{align}
	\sum_i H(\cK_i,\cK'_{i-1}\mid\cK_{i-1})
	\label{eq: sum cK}
\end{align}
over all rounds.
Later, \Cref{claim: sum cK} bounds the sum in \cref{eq: sum cK}, the accumulated structural information along the chain, in terms of the target size $\mathbb{E}[|\cE\setminus\cK|]$.

After the telescoping, the endpoint of the accounting is
\[
    H(F\mid \cK)\ge H(F)-\sum_{i=1}^t H(\cK_i,\cK_{i-1}\setminus\cK_i\mid \cK_{i-1}) -O(tp\log B).
\]
Equivalently, by the data processing inequality,
\[
	H(\cE)-H(\cE\mid\cK)=I(\cE:\cK)\le I(F:\cK)\le
	\sum_{i=1}^t H(\cK_i,\cK_{i-1}\setminus\cK_i\mid \cK_{i-1}) +O(tp\log B).
\]
This turns the probe-by-probe accounting into the main entropy inequality.

The encoding argument of Liu and Yu~\cite{liu2020lower} bounds $H(\cE\mid\cK)$ in terms of $\mathbb E[|\cE\setminus\cK|]$; \Cref{eq: H(cE|cK)} restates the form needed here.
Their paper does not isolate the lower bound on $H(\cE)$ in this form; rather, its proofs contain the distributional estimates from which it follows.
We extract the needed consequences in our notation (\Cref{lem_good_blocks,lem_entropy_lb,lem_large_gap1}) to lower bound $H(\cE)$ in \Cref{eq: size cE}.
Everything together forces $\cE\setminus\cK$ to be large, which in turn forces many hard queries to access the public memory cells.

\subsection{The inherited input distribution.}\label{sec: exact distribution}
We now introduce the distribution $\cD$ and the parameters of the \predz{} problem needed for the probe-by-probe information accounting in \Cref{sec_dist}.
The parameter choices for the RMQ reduction are given in \Cref{sec: reduction}.

The distribution is designed so that an RMQ structure can solve \predz{} without introducing a large entropy gap.
The RMQ answers are represented by a Cartesian tree, and the auxiliary integer $z$ encodes the tree structure not described by the answers to predecessor search.

For each $S_i=\{s^{(i)}_1<\cdots<s^{(i)}_u\}\subseteq [B]$ of the $d$ sub-databases, use sentinels $s^{(i)}_0=0$ and $s^{(i)}_{u+1}=B+1$ and define
\[
    W(S_i)=\prod_{j=0}^u C_{s^{(i)}_{j+1}-s^{(i)}_j-1}.
\]
Each Catalan factor counts the possible tree structures in a gap between successive datapoints.
The product therefore counts the residual tree structures once the tree path represented by $S_i$ is fixed.
The valid full inputs are tuples $(S_1,\ldots,S_d,z)$ satisfying
\[
    1\le z\le Z\cdot\prod_{i=1}^d W(S_i).
\]
The auxiliary integer $z$ encodes the residual choices, including the additional choices represented by the parameter $Z$.
In the RMQ reduction, $Z=\binom{2u}{u}^{r}$, where $r$ is the RMQ redundancy; the accounting itself allows general $Z$.

The distribution $\cD$ is uniform over these full inputs.
Equivalently, the sets $S_i$ are mutually independent, with
\[
    \Pr[S_i=S]=
    \frac{W(S)}{\displaystyle\sum_{\substack{S'\subseteq[B]\\|S'|=u}}W(S')},
\]
and, conditioned on them, $z$ is uniform over its allowed range.
This range compensates for the non-uniform set weights, so every full input has the same probability.
The normalizing sum over all $u$-element sets is a $(u+1)$-fold Catalan convolution, evaluated by Catalan's convolution formula~\cite{regev2012proof}.
Consequently, the full input entropy is
\begin{equation}
    H_{d,u,B,Z}
    =d\cdot\log\!\left(\sum_{\substack{S\subseteq[B]\\|S|=u}}W(S)\right)+\log Z.
    \label{eq: predz entropy}
\end{equation}
For $u=\Theta(\sqrt B)$, Liu and Yu~\cite[Lemma 2]{liu2020lower} evaluate this as $d(2B-u-\Theta(\log B))+\log Z$.
Under the parameter choices in \Cref{sec: reduction}, this is $2n-O(d\log B)$, close to the RMQ entropy $2n-1.5\log n+O(1)$.

\subsection{Public queries, certificates, and footprints.}\label{sec_qpub}
The public-query construction is inherited from Liu and Yu~\cite{liu2020lower}.
Its purpose is to expose high-entropy interval information at a controlled publishing cost.

\subsubsection{Selecting the public queries.}
Partition the subdatabases into $p$ consecutive blocks, each containing $m=ud/p$ datapoints.
Then choose regularly spaced predecessor queries within each \emph{good block}, defined in \Cref{def: good blocks}.
Writing $S_i=\{s^{(i)}_1<\cdots<s^{(i)}_u\}$, let
\[
	S^{(i)}_{\pt}\triangleq\left\{s^{(i)}_{m},s^{(i)}_{2m},s^{(i)}_{3m},\ldots,s^{(i)}_u\right\}
\]
be the set of elements used to partition each predecessor set into consecutive disjoint blocks containing exactly $m$ elements.
Write $S_{\pt}=(S^{(1)}_{\pt},\ldots,S^{(d)}_{\pt})$ for the collection of partition points.
The blocks in $S_i$ are $[s^{(i)}_{\ell m}+1,s^{(i)}_{(\ell+1)m}]$ for $0\le\ell<p/d$, with $s^{(i)}_0=0$.
Note that the blocks may have different lengths since the elements are randomly spread over the whole $[B]$.
We focus on good blocks, whose lengths are $\approx m^2$:
\begin{align}
	\cS_{\good}\triangleq
    \left\{(i,[x,y]):[x,y]\text{ is a block in }S_i,\quad
    \tfrac12 m^2\le y-x\le 2m^2\right\}.
	\label{def: good blocks}
\end{align}
A good block reproduces the square-root scale of a full sub-database: it has $m$ datapoints in a range of length $\Theta(m^2)$, just as a full sub-database has $u$ datapoints in $[B]$ with $u^2=\Theta(B)$.
This scale is motivated by the predecessor sets induced by uniformly random Cartesian trees, as explained in \Cref{sec: inherited estimates}.
\Cref{lem_good_blocks} shows that a constant fraction of the blocks are good in expectation.

Finally, we select $\Theta(\log^4 B)$ regularly spaced predecessor queries within each good block.
Recall that a block contains $m\triangleq ud/p$ elements and we assume $p<du/\log^4B$, so this selection is always possible.
Let $L\triangleq m^2/\log^4 B$ be the spacing, choose $\Delta$ uniformly from $[L]$, and set
\[
	Q_{\pub}\triangleq\bigcup_{(i,[x, y])\in \cS_{\good}}\left\{(i,x+j\cdot L+\Delta):j\geq 1,x+j\cdot L+\Delta<y\right\}.
\]
Note that there are at most $p$ good blocks, each good block has length at most $2L\log^4B=2m^2$, and we select $O(\log^4 B)$ queries from each block; hence $|Q_{\pub}|=O(p\log^4B)$.

Two adjacent queries delimit an interval whose non-emptiness is revealed by their answers.
For two consecutive selected positions $x+jL+\Delta$ and $x+(j+1)L+\Delta$ in a good block, define the occupancy bit
\[
	E^{(i,[x, y])}_j\triangleq\mathbf{1}_{\pred\left(i, x+jL+\Delta\right)\neq \pred\left(i, x+(j+1)L+\Delta\right)},
\]
which is $1$ precisely when the interval $(x+jL+\Delta,x+(j+1)L+\Delta]$ contains a datapoint.
Only intervals between consecutive selected public queries are considered below, and intervals are identified together with their subdatabase and block.
We let $\cE$ denote the set of these non-empty intervals between consecutive public queries in the good blocks.

\subsubsection{Memory accesses and certificates.}
We now formalize the sets of memory-cell addresses accessed by queries that certify the non-emptiness information in $\cE$.
For a query $q$, let $\probe(q)\subseteq[s]$ be the set of addresses of the memory cells accessed to answer $q$, where $[s]$ is the set of memory-cell addresses.
Note that $\probe(q)$ is a random variable since it depends on the database.
For $l\in[t]$, let $\probe_l(q)$ be the singleton containing the address of the memory cell read at the $l$-th memory access when answering $q$, or the empty set if the query has already terminated, and let
\[
	\probe_{\le l}(q)\triangleq\bigcup_{i\le l}\probe_i(q),\qquad
	\probe_{<l}(q)\triangleq\bigcup_{i< l}\probe_i(q),\qquad
	\probe_0(q)\triangleq\emptyset .
\]
For any set of queries $Q$, we define
\begin{align*}
	\probe(Q)&\triangleq\bigcup_{q\in Q}\probe(q),&
	\probe_0(Q)&\triangleq\emptyset,&
	\probe_l(Q)&\triangleq\bigcup_{q\in Q}\probe_l(q),\\
	\probe_{\le l}(Q)&\triangleq\bigcup_{q\in Q}\probe_{\le l}(q),&
	\probe_{<l}(Q)&\triangleq\bigcup_{q\in Q}\probe_{<l}(q),&
	\probe_{-l}(Q)&\triangleq [s]\setminus\probe_{\le l}(Q).
\end{align*}

A predecessor answer lying in an interval certifies that the interval is non-empty.
Let $\cK\subseteq\cE$ consist of the intervals $I$ for which some predecessor query $q$ returns a datapoint in $I$ and satisfies $\probe(q)\cap\probe(Q_{\pub})=\emptyset$.
The certifying query need not itself be a hard query: its returned datapoint identifies a hard query.

\subsubsection{Footprints and surviving certificates.}
We adopt the notion of \emph{footprint} from \cite{PV10}.
Unlike $\probe(q)$, which is an address set, a footprint is the sequence of contents read from the accessed memory cells, not their addresses.
We assume without loss of generality that an individual query never accesses the same memory-cell address twice.
For a query $q$, let $\foot_i(q)$ denote the $w$-bit word read at its $i$-th memory access, or the empty string $\varepsilon$ if the query has already terminated.
We do not pad executions that terminate early.
We define the prefixes of a footprint by
\begin{align*}
	\foot_{\le i}(q)&\triangleq \foot_1(q)\cdot\foot_2(q)\cdot\ldots\cdot\foot_i(q)
	\in(\HammingCube{w})^{|\probe_{\le i}(q)|},\\
	\foot_{<i}(q)&\triangleq \foot_{\le i-1}(q).
\end{align*}
We set $\foot(q)\triangleq \foot_{\le t}(q)$ and $\foot_{\le0}(q)=\foot_0(q)=\varepsilon$.
For a set of queries $Q=\{q_1<\cdots<q_k\}$ in lexicographical order, we define
\begin{align*}
	\foot_i(Q)&\triangleq \foot_i(q_1)\cdot\foot_i(q_2)\cdot\ldots\cdot\foot_i(q_k),\\
	\foot_{\le i}(Q)&\triangleq \foot_{\le i}(q_1)\cdot\foot_{\le i}(q_2)\cdot\ldots\cdot\foot_{\le i}(q_k),\\
	\foot_{<i}(Q)&\triangleq \foot_{\le i-1}(Q),\\
	\foot(Q)&\triangleq \foot(q_1)\cdot\foot(q_2)\cdot\ldots\cdot\foot(q_k)
		=\foot_{\le t}(Q).
\end{align*}
We set $\foot_{\le0}(Q)=\foot_0(Q)=\varepsilon$.
Thus $\foot_{\le i}(Q)$ contains $\sum_{q\in Q}|\probe_{\le i}(q)|$ words.
The same address may occur in several query footprints; we keep these repetitions for notational convenience.
For the rest of the proof of \Cref{lemma: unknown intervals}, abbreviate
\[
	F_i\triangleq \foot_i(Q_{\pub}),\qquad
	F_{\le i}\triangleq \foot_{\le i}(Q_{\pub}),\qquad
	F_{<i}\triangleq \foot_{<i}(Q_{\pub}),\qquad
	F\triangleq \foot(Q_{\pub}).
\]
Given $Q_{\pub}$ and the published bits, simulating each query $q$ identifies where $\foot_{<i}(q)$ ends, even if it terminates early.
Hence $F_{\le i}$ and the pair $(F_{<i},F_i)$ determine each another.
Observe that $F_{<i}$ determines $\probe_{\le i}(Q_{\pub})$ and therefore also $\probe_{-i}(Q_{\pub})$, given $Q_{\pub}$ and the published bits.
Finally, let
\[
	\rest_i(Q_{\pub})\in(\HammingCube{w})^{|\probe_{-i}(Q_{\pub})|}
\]
denote the binary string obtained by concatenating the $w$-bit contents of the memory cells at addresses in $\probe_{-i}(Q_{\pub})$, in increasing address order.

Let $\cK_i$ contain the intervals $I\in\cE$ for which some predecessor query $q$ returns a datapoint in $I$ and satisfies $\probe(q)\cap\probe_{\le i}(Q_{\pub})=\emptyset$.
Intuitively, $\cK_i$ is the set of non-empty intervals whose non-emptiness can still be certified when the memory cells at addresses in $\probe_{\le i}(Q_{\pub})$ are made unavailable.
We refer to this as deleting these memory cells; it does not change the addresses or contents of the remaining memory cells.
One can test these certificates using only the contents of the remaining memory cells: simulate each query and stop if it attempts to access an address in $\probe_{\le i}(Q_{\pub})$.
\Cref{algo: cKi} describes the resulting set.
Observe that $\cK_i$ is determined by $Q_{\pub}$, $F_{<i}$, and $\rest_i(Q_{\pub})$, given the published bits:
the first two pieces determine the unavailable address set $\probe_{\le i}(Q_{\pub})$, and $\rest_i(Q_{\pub})$ gives the contents of the memory cells at all remaining addresses.
Also note that the sequence is decreasing, with $\cK_0=\cE$ and $\cK_t=\cK$: $\cK_0\supseteq\cK_1\supseteq\cdots\supseteq\cK_t$.
For $i=0,\ldots,t-1$, define
\[
	\cK_i'\triangleq \cK_{i}\setminus \cK_{i+1},
\]
be the intervals that lose their last remaining certificates when the public memory cells accessed in the next memory-access round are deleted.

\begin{algorithm2e}
	\caption{Algorithm to compute $\cK_i$}
	\label{algo: cKi}
	$\cK_i\gets\emptyset$\;
	\ForEach{good block $(j,[x,y])$}
	{
		\For{$k\in[x,y]$}
		{
			\If{$\probe(\pred(j,k))\cap\probe_{\le i}(Q_{\pub})=\emptyset$ and $\pred(j,k)$ lies in an interval $I$ induced by $Q_{\pub}$ in this block}
			{
				$\cK_i\gets \cK_i\cup\{I\}$\;
			}
		}
	}
	\Return{$\cK_i$}\;
\end{algorithm2e}

\subsection{Conditioning convention.}\label{sec: conditioning}
In the accounting, we condition entropy and mutual-information expressions on the $p$ published bits, $S_{\pt}$, and $\Delta$, which together determine $Q_{\pub}$.
We suppress this common side information from the notation.
Describing it costs $O(p\log B)$ bits, as in the public-query construction of Liu and Yu; this cost is included in the local compression estimate.
The three block-distribution lemmas below describe the input distribution before conditioning on the algorithm's published bits; we explain their use under the accounting convention immediately afterwards.

\subsection{Inherited distributional and encoding estimates.}\label{sec: inherited estimates}
Two scaling observations explain the choice of good blocks and why the public-query answers are informative.
\begin{itemize}
	\item \emph{The square-root scale.}
Consider RMQ answers determined by a uniformly random Cartesian tree on $n$ positions.
Fix the left query endpoint at $\lfloor n/2\rfloor$, and let $S$ be the set of distinct answers as the right endpoint varies.
Each such query returns the predecessor of its right endpoint in $S$, and $S$ has roughly $\sqrt n$ elements~\cite{liu2020lower}.
This motivates the choice $u=\Theta(\sqrt B)$ for the hard distribution $\cD$.
In $\cD$, the number of datapoints is fixed at $u$, and their locations are weighted by $W(S_i)$, to simulate an unconditioned uniformly random Cartesian tree.
A good block keeps this scale locally, with $m$ datapoints in a range of length $\Theta(m^2)$.
\item \emph{Information at the public-query resolution.}
Given a uniform random RMQ database over $[n]$, if we divide $[n]$ into $c>1$ intervals of equal lengths, the number of non-empty intervals is likely $\approx\sqrt c$, and the collection of non-empty intervals has entropy $\approx\sqrt c$.
Thus, we can divide each good block into $\log^4 B$ intervals to obtain $\approx\log^2B$ non-empty intervals whose entropy is $\approx\log^2B$ bits.
\end{itemize}
The Catalan convolution formula~\cite{regev2012proof} underlies the counting calculations in Liu and Yu~\cite{liu2020lower} that yield the formal distributional estimates below.

The estimates collected here are inherited from Liu and Yu~\cite{liu2020lower}.
They describe the hard distribution and the cost of encoding missing intervals; the next subsection supplies the new relation between the public footprint and the surviving certificates.
\Cref{tab: inherited lemmas} collects their provenance and purpose.
The distributional statements are consequences in our notation, sometimes implicit in the cited proofs, rather than verbatim restatements.

\begin{table}[ht]
\centering
\begin{tabularx}{\linewidth}{@{}l l X@{}}
\toprule
Statement here & Source in \cite{liu2020lower} & Role \\
\midrule
\Cref{lem_good_blocks} & Lemma 6 & Many blocks are \emph{good blocks}. \\
\Cref{lem_entropy_lb} & Lemma 5 & Non-empty intervals in a good block have large entropy. \\
\Cref{lem_large_gap1} & Lemma 17 & Bounds the number of non-empty intervals in each good block. \\
\Cref{eq: size cE} & Lemmas 5, 6, 17 & Bounds the entropy and expected size of $\cE$. \\
\Cref{lem_enc_len}, \Cref{eq: H(cE|cK)} & Lemma 7 & Bounds $H(\cE\mid\cK)$ in terms of $|\cE\setminus\cK|$. \\
\bottomrule
\end{tabularx}
\caption{The inherited estimates used in the main lemma and its conversion into memory-access savings. The predecessor entropy formula and RMQ reduction are recalled separately in \Cref{sec: exact distribution,sec: reduction}.}
\label{tab: inherited lemmas}
\end{table}

The first estimate ensures $\Theta(p)$ good blocks in expectation.
The next two supply the entropy and expected number of non-empty intervals within a good block.
\begin{lemma}[Consequence of {\cite[Lemma 6]{liu2020lower}}]\label{lem_good_blocks}
For all integers $i\in[d]$ and $l\in[p/3d,2p/3d]$, we have
	\[
		\Pr\left[\text{block }\left(i,\left[s^{(i)}_{lm}+1, s^{(i)}_{(l+1)m}\right]\right)\text{ is good}\right]= \Omega(1).
	\]
\end{lemma}
\begin{lemma}[Consequence of {\cite[Lemma 5]{liu2020lower}}]\label{lem_entropy_lb}
	In a good block $(i,[x,y])$, let $\cE_{i,[x,y]}$ denote the set of non-empty intervals inside the good block.
	Then
	\[
		H(\cE_{i,[x,y]})= \Omega(\log^2 B).
	\]
	Moreover,
		for any good block $(i,[x,y])$, we have $\mathbb{E}[|\cE_{i,[x,y]}|]=\Omega(\log^2B)$.
\end{lemma}
\begin{lemma}[Consequence of {\cite[Lemma 17]{liu2020lower}}]\label{lem_large_gap1}
	In a good block $(i,[x,y])$, $\mathbb{E}[|\cE_{i,[x,y]}|]=O(\log^2B)$.
		\end{lemma}
Combining these estimates over the at most $p$ blocks gives the following bound, in the form needed for our accounting.
Conditioned on $S_{\pt}$, the datapoints in different blocks are independent~\cite[Claim 8]{liu2020lower}, so their conditional entropy contributions can be combined after the shift $\Delta$ is given.
Revealing the algorithm's $p$ published bits reduces the total entropy by at most $p$, leaving the $\Omega(p\log^2 B)$ bound unchanged in order.
The following claim therefore uses the conditioning convention of \Cref{sec: conditioning}.
\begin{restatable}{claim}{eqsizecE}
	\label{eq: size cE}
$H(\cE)=\Omega(p\log^2B)$ and $\mathbb{E}[|\cE|]=\Theta(p\log^2B)$.
\end{restatable}

The next lemma encodes $\cE$ when $\cK$ is already known, and thus upper bounds $H(\cE\mid\cK)$.
Its proof follows Liu and Yu's encoding almost identically and is included in \Cref{section: lem_enc_lem}.
\begin{lemma}[Adapted from {\cite[Lemma 7]{liu2020lower}}]\label{lem_enc_len}
	There is a prefix-free binary string $\cE({\cK})$ such that $\cE({\cK})$ and $\cK,Q_{\pub}$ together determine $\cE$. 	Moreover, we have the following bound on the length of $\cE({\cK})$:
	\[
		\mathbb{E}[\left|\cE({\cK})\right|]=O(\mathbb{E}[|\cE\setminus\cK|]\log(\mathbb{E}[|\cE|]/\mathbb{E}[|\cE\setminus\cK|])+p(\log\log B)^2).
	\]
\end{lemma}
Since this code together with $\cK$ determines $\cE$, its expected length bounds $H(\cE\mid\cK)$.
Thus we obtain the following immediate consequence.
\begin{restatable}{claim}{eqHcEcK}
	\label{eq: H(cE|cK)}
	$H(\cE|\cK)=O(\mathbb{E}[|\cE\setminus\cK|]\log(\mathbb{E}[|\cE|]/\mathbb{E}[|\cE\setminus\cK|])+p(\log\log B)^2)$.
\end{restatable}

\subsection{Proof of the main lemma.}\label{sec_dist}
\subsubsection{Local entropy claims.}\label{sec: local entropy}
First, conditioning on $\cK_i$ cannot substantially reduce the entropy of the $i$-th footprint layer $F_i$ once $F_{<i}$ is known.
The reason is that $\cK_i$ is computable from $F_{<i}$ and the string $\rest_i(Q_{\pub})$ of memory-cell contents at addresses outside $\probe_{\le i}(Q_{\pub})$.
Thus, if $\cK_i$ predicts $F_i$ much better than $F_{<i}$ alone does, then these remaining memory-cell contents would give a shorter-than-entropy encoding of $F_i$, the contents read in the $i$-th public memory-access round, and hence of the whole database.
\begin{claim}\label{claim: tail extent}	For any $0< i\le t$, 
	\begin{align*}
		H(F_i|F_{<i},\cK_i)\ge H(F_i|F_{<i})-O(p\log B).
	\end{align*}
\end{claim}
\begin{proof}[Proof of \Cref{claim: tail extent}]
	Note that
	\[
		H(F_i|F_{<i},\cK_i)
		\ge H(F_i|F_{<i},\rest_i(Q_{\pub}))
	\]
	since $\cK_i$ is known from $F_{<i}$ together with $\rest_i(Q_{\pub}),Q_{\pub}$.
	Thus it suffices to prove 
	\begin{align}
	H(F_i|F_{<i},\rest_i(Q_{\pub}))
	\ge H(F_i|F_{<i}) -O(p\log B).\label{eq: encode argument}
	\end{align}

	Fix $i\in[t]$.
	Consider the following encoding scheme for the database:
\begin{enumerate}[label=({\arabic*})]
	\item
		write down the $p$ published bits;
	\item
		write down $S_{\pt}$;
	\item
		sample $\Delta\in[L]$ uniformly at random, write down $\Delta$;
	\item
		write down the contents of the memory cells at addresses in $\probe_{1}(Q_{\pub}),\dots,\probe_{i-1}(Q_{\pub})$.
		For each $i$, process the addresses in increasing order and skip those memory cell whose contents have already been written, so this step uses $w\cdot|\probe_{<i}(Q_{\pub})|$ bits;
	\item
		write down $\rest_i(Q_{\pub})$, the contents of the memory cells at addresses in $\probe_{-i}(Q_{\pub})$, in increasing address order, using $w\cdot|\probe_{-i}(Q_{\pub})|=|\rest_i(Q_{\pub})|$ bits;
	\item
		encode the contents of the memory cells at addresses in $\probe_i(Q_{\pub})\setminus\probe_{<i}(Q_{\pub})$, using an optimal conditional prefix-free code of expected length $H(F_i\mid\rest_i(Q_{\pub}),F_{<i})+O(1)$ bits.
\end{enumerate}
	Recall that $Q_{\pub}$ is known from $S_{\pt}$ together with $\Delta$.
	The decoder simulates the public queries round by round and recovers the previously unseen memory-cell addresses and contents from step (4).
	This reconstructs $F_{<i}$ and determines the addresses accessed in round $i$.
	Step (5) supplies the memory-cell contents at addresses outside $\probe_{\le i}(Q_{\pub})$, and step (6) supplies the contents at addresses accessed in round $i$.
	Thus the codeword recovers the whole data structure, including its published bits.
	Furthermore, we can decode the input database from the data structure, thus the codeword must be of length at least $H_{d,u,B,Z}$ bits in expectation.

	Obviously the steps (1)--(3) consume $O(p\log B)$ bits.
	The encoding scheme partitions the memory-cell address space $[s]$ into three disjoint address sets:
	(i) $\probe_{<i}(Q_{\pub})$, (ii) $\probe_i(Q_{\pub})\setminus\probe_{<i}(Q_{\pub})$, and (iii) $\probe_{-i}(Q_{\pub})$.
	Given $Q_{\pub}$ and the published bits, $F_{<i}$ determines this partition, including the addresses accessed in memory-access round $i$.
	The data structure represents the contents of the memory cells at addresses in $\probe_i(Q_{\pub})\setminus\probe_{<i}(Q_{\pub})$ using
	\[
		w\cdot\mathbb{E}[|\probe_i(Q_{\pub})\setminus\probe_{<i}(Q_{\pub})|]
		\ge H(F_i\mid F_{<i})
	\]
	bits in expectation.
	On the other hand, the encoding scheme represents the same memory-cell contents using $H(F_i\mid\rest_i(Q_{\pub}),F_{<i})+O(1)$ bits in expectation, at the cost of an extra $O(p\log B)$-bit description.
	If $H(F_i\mid F_{<i})-H(F_i\mid\rest_i(Q_{\pub}),F_{<i})=\omega(p\log B)$, the last step would save $\omega(p\log B)$ bits in writing down the contents of the memory cells at addresses in $\probe_i(Q_{\pub})\setminus\probe_{<i}(Q_{\pub})$.
	This is impossible, since we do not waste any bit in writing down the contents of the memory cells at addresses in $\probe_{<i}(Q_{\pub})$ and $\probe_{-i}(Q_{\pub})$.
	In other words, if \Cref{eq: encode argument} does not hold, the codeword is of length $H_{d,u,B,Z}-\omega(p\log B)$ bits in expectation, a contradiction.
	Therefore, \Cref{eq: encode argument} holds.
\end{proof}

Second, when we replace the condition $\cK_i$ by $\cK_{i-1}$, the only extra information that may be lost is the partition of $\cK_{i-1}$ into intervals that survive the deletion of round $i$ and intervals that do not.
This is the price for making the previous display telescope from $H(F_{\le i}\mid\cK_i)$ to $H(F_{\le {i-1}}\mid\cK_{i-1})$.
Equivalently, this claim accounts for the information needed to specify how the certifiable set changes in each memory-access round via examining the nested chain
$\cK_t\subseteq\cK_{t-1}\subseteq\cdots\subseteq\cK_1\subseteq\cK_0$
one layer at a time.
The entropy terms accumulated here will later be converted, in \Cref{claim: sum cK}, into the target quantity $\mathbb{E}[|\cE\setminus\cK|]$.
\begin{claim}\label{claim: condition manipulate}	For any $0< i\le t$, 
	\begin{align*}
		H(F_{<i}|\cK_i)\ge H(F_{<i}|\cK_{i-1})-H(\cK_i,\cK'_{i-1}|\cK_{i-1}).
	\end{align*}
\end{claim}
\begin{proof}[Proof of \Cref{claim: condition manipulate}]
	Note that $H(F_{<i}\mid \cK_i)\ge H(F_{<i}\mid \cK_i,\cK_{i-1},\cK'_{i-1})$.
	Moreover,
	\begin{align*}
	& H(F_{<i}\mid \cK_{i-1})
	-H(F_{<i}\mid \cK_i,\cK_{i-1},\cK'_{i-1})\\
	=& I(F_{<i}:\cK_{i},\cK'_{i-1}\mid \cK_{i-1})
	\le H(\cK_{i},\cK'_{i-1}\mid \cK_{i-1}).
	\end{align*}
	Rearranging gives the claim.
\end{proof}

\subsubsection{Telescoping the accounting.}
We now telescope the two local estimates to prove the main lemma~\Cref{lemma: unknown intervals}.
\begin{proof}[Proof of \Cref{lemma: unknown intervals}]
By the chain rule and the two claims, for every $i\in[t]$,
\begin{align*}
	H(F_{\le i}|\cK_i)
	&=H(F_i|F_{<i},\cK_i)+H(F_{<i}|\cK_i)\\
	&\ge H(F_i|F_{<i})+H(F_{<i}|\cK_{i-1})-O(p\log B)-H(\cK_i,\cK'_{i-1}|\cK_{i-1}).
\end{align*}
Iterating this inequality from $i=t$ to $i=1$ and using $\cK_t=\cK$, $\cK_0=\cE$, and $F_{\le t}=F$, we obtain
\begin{align}
	H(F|\cK)
	\ge H(F)-O(tp\log B)-\sum_{i=1}^tH(\cK_i,\cK'_{i-1}|\cK_{i-1}).
	\label{eq: examine algo}
\end{align}
Equivalently,
\begin{align}
	I(F:\cK)\le O(tp\log B)+\sum_{i=1}^tH(\cK_i,\cK'_{i-1}|\cK_{i-1}).
	\label{eq: footprint-info}
\end{align}
This is the probe-by-probe accounting inequality.
The mutual information $I(F:\cK)$ measures how much of the footprint's information is shared with the intervals that remain certifiable without the public memory cells.
The inequality bounds this shared information by the intermediate changes $\cK_{i-1}\to\cK_i$, plus the description cost:
at round $i$, conditioned on $\cK_{i-1}$, it suffices to specify which intervals survive in $\cK_i$ and which intervals disappear into $\cK'_{i-1}$.
Thus the proof does not choose a single important access round.
Instead, it charges the mutual information between $\cK$ and $F$ to the sum of all local changes along the query algorithm's adaptive memory-access sequence.

The set $\cE$ is determined by the answers to $Q_{\pub}$, and hence by $F$.
By the data processing inequality, $I(\cE:\cK)\le I(F:\cK)$.
Adding $H(\cE|\cK)$ to both sides gives the main inequality
\begin{align}
	\sum_{i=1}^tH(\cK_i,\cK'_{i-1}|\cK_{i-1})+H(\cE|\cK)+O(tp\log B)\ge H(\cE).
	\label{eq: goal}
\end{align}
The summation can be restated in terms of $\mathbb E[|\cE\setminus\cK|]$ in \Cref{claim: sum cK}.
The conditional entropy $H(\cE\mid\cK)$ is bounded by the encoding argument of \cite{liu2020lower} in terms of the desired $\mathbb E[|\cE\setminus\cK|]$, restated here in the form needed for our accounting in \Cref{eq: H(cE|cK)}.
Since $H(\cE)$ is large, this forces $\mathbb E[|\cE\setminus\cK|]$ to be large.

For the accumulated changes in the chain, we use the following estimate, proved immediately after the main lemma.
\begin{restatable}{claim}{claimsumcK}
	\label{claim: sum cK}
	$\sum_{i=1}^tH(\cK_i,\cK'_{i-1}|\cK_{i-1})=O(\mathbb{E}[|\cE\setminus\cK|]\log(t\cdot\mathbb{E}[|\cE|]/\mathbb{E}[|\cE\setminus\cK|]))$.
\end{restatable}
Let $M\triangleq \E[|\cE\setminus\cK|]$ and $A\triangleq \E[|\cE|]=\Theta(p\log^2B)$.
Plugging \Cref{claim: sum cK,eq: H(cE|cK),eq: size cE} into \Cref{eq: goal} gives
\begin{align*}
	M\log(tA/M)+M\log(A/M)+tp\log B+p(\log\log B)^2=\Omega(A).
\end{align*}
Since we are in the regime $t=o(\log B)$, the last two additive terms are $o(A)$.
Moreover, $M\le A$ by definition, and the first two terms are dominated by $O(M\log(tA/M))$.
If $M=o(A/\log t)$, then $M\log(tA/M)=o(A)$, resulting in a contradiction.
Therefore $M=\Omega(A/\log t)=\Omega(p\log^2B/\log t)$, as required.
\end{proof}

We will use the following elementary averaging inequality to prove \Cref{claim: sum cK}.
\begin{fact}\label{fact: inequality}
	For any two jointly distributed positive random variables $X,Y$ with finite support,
	\[
		\mathbb{E}[X\log(Y/X)]\le
		\mathbb{E}[X]\log(\mathbb{E}[Y]/\mathbb{E}[X]).
	\]
\end{fact}
\begin{proof}
	\begin{align*}
		&\mathbb{E}[X\log(Y/X)]\\
		=&\mathbb{E}[Y(X/Y)\log(Y/X)]\\
		=&\sum_y\Pr[Y=y]\cdot y\cdot\mathbb{E}[(X/Y)\log(Y/X)\mid Y=y]\\
		\intertext{By the concavity of $x\log(1/x)$,}
		\le&\sum_y\Pr[Y=y]\cdot y\cdot\mathbb{E}[X/Y\mid Y=y]\log(1/\mathbb{E}[X/Y\mid Y=y])\\
		=&\mathbb{E}[Y] \sum_y(\Pr[Y=y]\cdot y/\mathbb{E}[Y])\mathbb{E}[X/Y\mid Y=y]\log(1/\mathbb{E}[X/Y\mid Y=y])
		\intertext{Observe that the weights $\Pr[Y=y]\cdot y/\mathbb{E}[Y]$ form a probability distribution on $y$; denote it by $\mu$.}
		=&\mathbb{E}[Y]\mathbb{E}_{y\sim\mu}[\mathbb{E}[X/Y\mid Y=y]\log(1/\mathbb{E}[X/Y\mid Y=y])]
		\intertext{Since $\mathbb{E}_{y\sim\mu}[\mathbb{E}[X/Y\mid Y=y]]=\mathbb{E}[X]/\mathbb{E}[Y]$, another application of the concavity of $x\log(1/x)$ gives}
		\le&\mathbb{E}[X]\log(\mathbb{E}[Y]/\mathbb{E}[X]).
	\end{align*}
\end{proof}

It remains to prove \Cref{claim: sum cK}, which converts the information in the certificate chain into the unknown-interval count.
\claimsumcK*
\begin{proof}
Observe that $(\cK_{i},\cK'_{i-1})$ is a partition of $\cK_{i-1}$, thus 
\[
	H(\cK_i,\cK'_{i-1}|\cK_{i-1})=O\left(\mathbb{E}\left[\log\binom{|\cK_{i-1}|}{|\cK'_{i-1}|}\right]\right)=O\left(\mathbb{E}[|\cK'_{i-1}|\log(|\cK_{i-1}|/|\cK'_{i-1})]\right).
\]
Therefore, it suffices to prove
\begin{align*}
	&\sum_{i\in[t]}\mathbb{E}[|\cK'_{i-1}|\log(|\cK_{i-1}|/|\cK'_{i-1}|)]\\
		\le&\sum_{i\in[t]}\mathbb{E}[|\cK'_{i-1}|]\log(\mathbb{E}[|\cK_{i-1}|]/\mathbb{E}[|\cK'_{i-1}|])&(\text{\Cref{fact: inequality}})\\
	=&t\cdot\mathbb{E}_{i\sim[t]}\left[\mathbb{E}[|\cK'_{i-1}|]\log(\mathbb{E}[|\cK_{i-1}|]/\mathbb{E}[|\cK'_{i-1}|])\right]\\
	\intertext{Observe that $\mathbb{E}_{i\sim[t]}[\mathbb{E}[|\cK'_{i-1}|]]=\mathbb{E}[|\cE\setminus\cK|]/t$ and $\mathbb{E}_{i\sim[t]}[\mathbb{E}[|\cK_{i-1}|]]\le\mathbb{E}[|\cE|]$. By \Cref{fact: inequality} again,}
	\le&\mathbb{E}[|\cE\setminus\cK|]\log(t\cdot\mathbb{E}[|\cE|]/\mathbb{E}[|\cE\setminus\cK|]),
\end{align*}
where we let $i\sim[t]$ denote that $i$ is sampled from $[t]$ uniformly at random.
\end{proof}

\section{Deriving the Lower Bounds from the Main Lemma.}\label{sec_lower}
Recall that $\cE\setminus\cK$ is the set of \emph{unknown intervals}~\cite{liu2020lower}: non-empty intervals whose non-emptiness cannot be certified without accessing the public memory cells.
By the definition of unknown intervals, every hard query at a datapoint in such an interval must access at least one public memory cell.
Publishing the addresses and contents of the public memory cells therefore saves at least one memory access for each such hard query.
We derive the main theorem from the lower bound on $\mathbb E[|\cE\setminus\cK|]$ in the following three steps.
\[
    \underbrace{\Cref{lemma: unknown intervals}}_{\text{unknown intervals}}
    \ \Longrightarrow\
    \underbrace{\Cref{lem_int}}_{\text{access savings}}
    \ \Longrightarrow\
    \underbrace{\Cref{lem_predz}}_{\text{\predz{} lower bound}}
    \ \Longrightarrow\
    \underbrace{\Cref{thm_main}}_{\text{RMQ lower bound}}.
\]
The first implication uses Liu and Yu's sparse-interval estimate~\cite{liu2020lower} to convert the unknown-interval count into datapoint counts, and then into memory-access savings.
The second and third follow the published-bit recursion and RMQ reduction of Liu and Yu~\cite{liu2020lower}.
Thus the improved saving below replaces their Lemma 4; it is the only change needed by that framework.

\subsection{From unknown intervals to memory-access savings.}\label{sec: efficient round-elimination}
\Cref{lemma: unknown intervals} lower-bounds the expected number of unknown intervals.
To obtain memory-access savings, we must also count the datapoints in these intervals, since each datapoint corresponds to a hard query.
The following inherited lemma bounds the expected number of sparse intervals.
Together with our unknown-interval bound, it shows that the unknown intervals collectively contain many datapoints in expectation.
\begin{lemma}[{\cite[Lemma 15]{liu2020lower}}]\label{lem_small_interval}
  For $l\leq m/\log^2B=O(\sqrt{L})$, in the good blocks, the expected number of intervals that have between $l/2$ and $l$ elements is at most $O(pl\log^4B/m+p)$.
\end{lemma}

Each block contains exactly $m=ud/p$ datapoints.
We use \Cref{lem_small_interval} to subtract sparse intervals from the unknown interval count, leaving enough datapoints to obtain the following overlap bound.
\begin{samepage}
\begin{lemma}[Efficient round-elimination, improved {\cite[Lemma 4]{liu2020lower}}]\label{lem_int}
	Given a \predz{} data structure with at most $H_{d,u,B,Z}$ bits of memory, $p$ published bits for $d\le p<du/\log^4 B$, and worst-case $\pred$ query time $t=o(\log B)$, there exists a random set $Q_{\pub}$ of $O(p\log^4 B)$ predecessor queries such that
	\[
		\E\left[\frac{1}{|\cQ|}\sum_{q\in \cQ} |\probe(q)\cap \probe(Q_{\pub})|\right]= \Omega\left(\frac{1}{\log^2 t}\right),
	\]
	where the expectation is taken over the random input $(\{S_1,\ldots,S_d\},z)\sim\cD$ and the choice of $Q_{\pub}$.
\end{lemma}
\end{samepage}

\begin{proof}[Proof of \Cref{lem_int}]
	Since $t=o(\log B)$, \Cref{lemma: unknown intervals} gives $\mathbb{E}[|\cE\setminus\cK|]=\Omega(p\log^2 B/\log t)$.
	Recall that $\cK'\subseteq\cE$ consist of intervals containing a datapoint whose hard query $q$ satisfies $\probe(q)\cap\probe(Q_{\pub})=\emptyset$.
	$\cK'$ is a stricter version of $\cK$, obtained by requiring the certifying query to be hard.
	Thus $\cE\setminus\cK'$ consists of intervals in which every hard query at a datapoint satisfies $\probe(q)\cap\probe(Q_{\pub})\ne\emptyset$.
	In other words, we save at least one memory access when answering each hard query at a datapoint in $\cE\setminus\cK'$ if the contents of the memory cells at addresses in $\probe(Q_{\pub})$ are available in the published bits.
	Since $\cK'\subseteq\cK\subseteq\cE$, we have $\mathbb{E}[|\cE\setminus\cK'|]\ge\mathbb{E}[|\cE\setminus\cK|]=\Omega(p\log^2B/\log t)$.

	Let $c_t>0$ be the constant in the lower bound of \Cref{lemma: unknown intervals}, and let $c_l>0$ be an upper-bound constant in \Cref{lem_small_interval}.
	By \Cref{lem_small_interval}, in the good blocks, the expected number of intervals that contain at most $l$ elements is at most
	\begin{align*}
		\sum_{i=0}^{\lfloor\log l\rfloor} c_l(p2^i\log^4 B/m+p)
        =O(pl\log^4 B/m+p\log l).
	\end{align*}
Choose $l=\Theta(m/(\log^2 B\log t))$ with a sufficiently small constant.
The first term is then at most a sufficiently small constant times $p\log^2 B/\log t$.
The second is $O(p\log B)=o(p\log^2 B/\log t)$ because $t=o(\log B)$.
Consequently, the expected number of intervals in $\cE\setminus\cK'$ containing more than $l$ datapoints is $\Omega(p\log^2 B/\log t)$.
Every hard query at a datapoint in each such interval accesses a memory cell at an address in $\probe(Q_{\pub})$.
	Thus the expected number of hard queries $q$ such that $\probe(q)\cap\probe(Q_{\pub})\ne\emptyset$ is at least $\Omega(pm/\log^2t)$.
	The lemma then follows from the fact that there are exactly $ud=pm$ hard queries.
\end{proof}

\subsection{The inherited publishing recursion.}\label{sec: section predz lower bound}
Publishing the address set $\probe(Q_{\pub})$ and the contents of the corresponding memory cells costs $O(pwt\log^4 B)$ bits and allows every query to skip its accesses to these memory cells.
\Cref{lem_int} therefore saves $\Omega(1/\log^2 t)$ expected memory accesses for an average hard query in each elimination step.
After $O(t\log^2 t)$ steps, continuing with the same guarantee would make the expected number of memory accesses negative.
The number of published bits must reach the range where the lemma no longer applies before that happens.
This gives the following bound.
\begin{lemma}[Improved {\cite[Lemma 3]{liu2020lower}}]\label{lem_predz}
For any parameters $d,u,B$ and $Z$ satisfying $u=\Theta(\sqrt{B})$, for any data structure that uses at most $H_{d,u,B,Z}+O(d\log B)$ bits of space and answers $\pred$ queries in time $t$ for solving \predz{} under distribution $\cD$, it must hold
	\begin{align}
		(wt\log B)^{O(t\log^2 t)}\geq B,\label{eq: pred lower bound}
	\end{align}
	in the cell-probe model with word-size $w$.
\end{lemma}
\begin{proof}[Proof sketch of \Cref{lem_predz}, following {\cite[Lemma 3]{liu2020lower}}]
Move the $O(d\log B)$ redundant bits into the published part, and let $p_0=O(d\log B)$ be the initial number of published bits.
It suffices to consider $t=o(\log B)$, since otherwise the claimed bound is trivial.
Keep $t$ as the original worst-case query bound throughout the recursion.

As long as $d\le p<du/\log^4 B$, apply \Cref{lem_int} and publish the addresses and contents of the public memory cells.
The number of published bits grows by at most a factor $O(wt\log^4 B)$, while the expected number of memory accesses for an average hard query decreases by at least $\Omega(1/\log^2 t)$.
The expectation is over $\cD$, a uniformly chosen hard query of the current database, and the choices of public queries.
This decrease cannot continue for more than $\omega(t\log^2 t)$ steps, since the initial query cost is at most $t$ and memory-access counts are nonnegative.
Hence, the number of published bits reaches $du/\log^4 B$ within that many steps, and
\[
    O(d\log B)\bigl(O(wt\log^4 B)\bigr)^{O(t\log^2 t)}
    \ge \frac{du}{\log^4 B}.
\]
Using $u=\Theta(\sqrt B)$ and absorbing the logarithmic factors gives
$(wt\log B)^{O(t\log^2 t)}\ge B$.
\end{proof}

\subsection{The inherited reduction to RMQ.}\label{sec: reduction}
For completeness, we recall the part of Liu and Yu's reduction~\cite[Theorem 1]{liu2020lower} needed to substitute our stronger \predz{} bound.
The distribution $\cD$ and its entropy $H_{d,u,B,Z}$ were defined in \Cref{sec: exact distribution}; the parameters are
\[
    d=2r,\qquad B=\lfloor n/d\rfloor-1,\qquad
    u=\lfloor\sqrt B\rfloor,\qquad Z=\binom{2u}{u}^{r}.
\]
Up to the usual integer rounding, $H_{d,u,B,Z}=2n-O(d\log B)$.
The reduction turns an RMQ structure with $2n-1.5\log n+r$ bits and query time $t$ into a \predz{} structure with $H_{d,u,B,Z}+O(d\log B)$ bits and predecessor query time $t$.
This is an inherited reduction; our change is the stronger constraint supplied by \Cref{lem_predz}.

\begin{proof}[Proof sketch of \Cref{thm_main}, following {\cite[Theorem 1]{liu2020lower}}]
The reduction embeds each predecessor set in a consecutive block of the RMQ array.
Within each embedded block, range-minimum queries with one endpoint fixed recover the corresponding predecessor answers.

The auxiliary integer $z$ encodes the remaining Cartesian-tree structure, keeping the input entropy of \predz{} within $O(d\log B)$ bits of the RMQ entropy.
Recall that 
\[
	z\in\left[Z\cdot\prod_{i=1}^d W(S_i)\right], \text{\quad and\quad}
 W(S_i)=\prod_{j=0}^u C_{s^{(i)}_{j+1}-s^{(i)}_j-1}.
\]
The components of the database for \predz{} play the following roles in the reduction:
\begin{enumerate}
	\item The $u$ datapoints $s^{(i)}_1,s^{(i)}_2,\ldots,s^{(i)}_u$ form a path from the root to a leaf in the Cartesian tree of $S_i$, where $s^{(i)}_j$ is the left child of $s^{(i)}_{j-1}$ for even $i$ and the right child for odd $i$; 
	\item The quotient $z/Z$ specifies the sub-tree structures in all gaps between consecutive datapoints. The factor $C_{s^{(i)}_{j+1}-s^{(i)}_j-1}$ is the number of possible Cartesian-tree structures of the right subtree of $s^{(i)}_j$ for even $i$ and the left subtree for odd $i$;
	\item The remainder $z\mod Z$ specifies the merge choices for the pairs $(S_1,S_2),(S_3,S_4),\ldots$. Each pair has $\binom{2u}{u}$ interleavings, giving $Z=\binom{2u}{u}^{r}$ choices in total.
\end{enumerate}
Thus the assumed RMQ structure gives the \predz{} structure just described.

Apply \Cref{lem_predz} to obtain
\[
    B\le (wt\log B)^{O(t\log^2 t)}.
\]
In the nontrivial range $r=o(n)$, we have $B=\Theta(n/r)$.
For $w=\Omega(\log n)$ and $t\le\log n$, the factor $wt\log B$ is $w^{O(1)}$, and hence
\[
    n/r\le w^{O(t\log^2 t)}.
\]
Rearranging proves the theorem.
For $t>\log n$, the asserted bound is weaker than the trivial constant-redundancy bound. %
\end{proof}

\section*{Acknowledgments.}
The author thanks Yi Li and Huacheng Yu for helpful discussions, and the anonymous reviewers for detailed comments that improved the exposition.
Part of this work was done while the author was a postdoctoral researcher at Nanyang Technological University, supported by Singapore Ministry of Education (AcRF) Tier 2 grant MOE2018-T2-1-013, and at BARC, University of Copenhagen, supported by VILLUM Foundation grant 16582.
The author is supported by the ``111 Center'' (No.~B26023) and the Fundamental and Interdisciplinary Disciplines Breakthrough Plan of the Ministry of Education of China (No.~JYB2025XDXM118).

\section*{AI Use Statement.}
AI (GPT-6 Astra) was used solely to assist with manuscript writing and language polishing.

\bibliography{refs}
\bibliographystyle{siamplain}

\appendix

\section[Proof of the encoding-length lemma]{Proof of \Cref{lem_enc_len}}\label{section: lem_enc_lem}
First note that $\cK\subseteq\cE$.
Hence it suffices to encode in $\cE(\cK)$, the set $\cE\setminus\cK$.
First, we present an encoding scheme which encodes the set $\cE_{(j,[x,y])}\setminus \cK_{(j,[x,y])}$ in a prefix-free binary string $\cE_{(j,[x,y])}\left(\cK_{\left(j,[x,y]\right)}\right)$ for some good block $(j,[x,y])$, where $\cE_{(j,[x,y])}$ is the set of non-empty intervals in the good block $(j,[x,y])$ and $\cK_{(j,[x,y])}\triangleq\cK\cap\cE_{(j,[x,y])}$.
Then we show that we can efficiently encode $\cE\setminus\cK$ block by block.

\subsection{An encoding scheme within a good block}
\subsubsection[Encoding within a good block]{Encoding $\cE_{(j,[x,y])}$ given $\cK_{(j,[x,y])},Q_{\pub}$}
	Let $K_{ne}\triangleq |\cE_{(j,[x,y])}|,K_{un}\triangleq |\cE_{(j,[x,y])}\setminus\cK_{(j,[x,y])}|$.
	Let $I_1,I_2,\ldots,I_{K_{ne}}\in\cE_{(j,[x,y])}$ be all the elements in the increasing order.
	Let $I_{i_1}, I_{i_2},\ldots,I_{i_{K_{un}}}\in \cE_{(j,[x,y])}\setminus\cK_{(j,[x,y])}$ be all the elements to be encoded, where $i_1<i_2<\ldots<i_{K_{un}}$.
	We first write down $K_{ne}$, then for $a=1,\ldots,K_{un}$, we do the following:
	\begin{enumerate}
	    \item write down $i_a-i_{a-1}$ ($i_0$ is assumed to be $0$);
	    \item write down $I_{i_a}-I_{i_a-1}$.
	\end{enumerate}
	All integers are encoded using the folklore prefix-free encoding which takes $O(\log N)$ bits to encode a non-negative integer $N$.
	This completes $\cE_{(j,[x,y])}(\cK_{(j,[x,y])})$.

\subsubsection[Decoding within a good block]{Decoding $\cE_{(j,[x,y])}$ given $\cE_{(j,[x,y])}(\cK_{(j,[x,y])})$ and $\cK_{(j,[x,y])},Q_{\pub}$}
	 We read $K_{ne}$ (which, together with $|\cK_{(j,[x,y])}|,Q_{\pub}$, determines $K_{un}$), and for $a=1,\ldots,K_{un}$, do the following:
	\begin{enumerate}
		\item read the next integer and recover $i_a$;
		\item for $i=i_{a-1}+1,\ldots,i_a-1$, let $I_i$ be the next element in $\cE_{(j,[x,y])}$ ($i_0$ is assumed to be $0$);
		\item read the next integer and recover $I_{i_a}$.
	\end{enumerate}
	Finally, for $i=i_{K_{un}}+1,\ldots,K_{ne}$, let $I_i$ be the next element in $\cE_{(j,[x,y])}$.
	This recovers all $I_1,\ldots,I_{K_{ne}}$, hence, decodes $\cE_{(j,[x,y])}$.

\subsubsection[Length of the block encoding]{The length of $\cE_{(j,[x,y])}(\cK_{(j,[x,y])})$}
	Next, we analyze the expected length of $\cE_{(j,[x,y])}(\cK_{(j,[x,y])})$.
	$K_{ne}$ takes $O(\log \log B)$ bits to encode.
	Then for $a=1,\ldots,K_{un}$, $(i_a-i_{a-1})$ takes $O(\log (i_a-i_{a-1}))$ bits to encode.
	Since all these integers sum up to (at most) $K_{ne}$, by concavity of $\log$, the total number of bits used to encode $\{i_a-i_{a-1}\}$ is at most 
	\[
		O(K_{un}\cdot \log\frac{K_{ne}}{K_{un}}).
	\]
			Then by \Cref{fact: inequality}, 	the expected encoding length of all $i_a-i_{a-1}$ is at most
	\begin{equation}\label{eqn_length_1}
        O\left(\mathbb{E}[|\cE_{(j,[x,y])}\setminus\cK_{(j,[x,y])}|]\log\frac{\mathbb{E}[|\cE_{(j,[x,y])}|]}{\mathbb{E}[|\cE_{(j,[x,y])}\setminus\cK_{(j,[x,y])}|]}\right)
			\end{equation}

	Next, the value $I_{i_a}-I_{i_a-1}$ takes $O(\log (I_{i_a}-I_{i_a-1}))$ bits to encode.
	For all $a$ such that $I_{i_a}-I_{i_a-1}\le \left(\frac{\mathbb{E}[|\cE_{(j,[x,y])}|]}{\mathbb{E}[|\cE_{(j,[x,y])}\setminus\cK_{(j,[x,y])}|]}\right)^2$, their total encoding length is at most
	\[
		O\left(K_{un}\cdot \log\frac{\mathbb{E}[|\cE_{(j,[x,y])}|]}{\mathbb{E}[|\cE_{(j,[x,y])}\setminus\cK_{(j,[x,y])}|]}\right),
	\]
	and its expectation is at most
	\begin{equation}\label{eqn_length_2}
        O\left(\mathbb{E}[|\cE_{(j,[x,y])}\setminus\cK_{(j,[x,y])}|]\log\frac{\mathbb{E}[|\cE_{(j,[x,y])}|]}{\mathbb{E}[|\cE_{(j,[x,y])}\setminus\cK_{(j,[x,y])}|]}\right)
	\end{equation}

    \begin{lemma}[{\cite[Lemma 17]{liu2020lower}}]\label{lem_large_gap}
		In a good block, for any $t\ge 1$, the expected number of non-empty interval pairs that have between $t-1$ and $2t$ empty intervals and no non-empty interval in between is at most $O\left(\sqrt{(\log^4B)/t}+1\right)$.
\end{lemma}
	For all $a$ such that $I_{i_a}-I_{i_a-1}> \left(\frac{\mathbb{E}[|\cE_{(j,[x,y])}|]}{\mathbb{E}[|\cE_{(j,[x,y])}\setminus\cK_{(j,[x,y])}|]}\right)^2$, by \Cref{lem_large_gap}, the summation of their expected encoding lengths is at most
	\begin{align*}
		&\ \sum_{t=2^b\cdot\left(\frac{\mathbb{E}[|\cE_{(j,[x,y])}|]}{\mathbb{E}[|\cE_{(j,[x,y])}\setminus\cK_{(j,[x,y])}|]}\right)^2:b\geq 0, t\le \log^4B} O\left(\left(\sqrt{(\log^4B)/t}+1\right)\log t\right)\nonumber\\
		\le& O\left((\log^2B)\mathbb{E}[|\cE_{(j,[x,y])}\setminus\cK_{(j,[x,y])}|]/\mathbb{E}[|\cE_{(j,[x,y])}|]\cdot\log \left(\frac{\mathbb{E}[|\cE_{(j,[x,y])}|]}{\mathbb{E}[|\cE_{(j,[x,y])}\setminus\cK_{(j,[x,y])}|]}\right)+\log^2\log B\right).
	\end{align*}
    Since $\mathbb{E}[|\cE_{(j,[x,y])}|]=\Omega(\log^2B)$ by \Cref{lem_entropy_lb}, it is at most
    \begin{align}
        O\left(\mathbb{E}[|\cE_{(j,[x,y])}\setminus\cK_{(j,[x,y])}|]\cdot\log \left(\frac{\mathbb{E}[|\cE_{(j,[x,y])}|]}{\mathbb{E}[|\cE_{(j,[x,y])}\setminus\cK_{(j,[x,y])}|]}\right)+\log^2\log B\right).\label{eqn_length_3}
    \end{align}

	Finally, summing up \Cref{eqn_length_1}, \Cref{eqn_length_2} and \Cref{eqn_length_3}, the expected length of $\cE_{(j,[x,y])}(\cK_{(j,[x,y])})$ is at most
	\[
        O\left(\mathbb{E}[|\cE_{(j,[x,y])}\setminus\cK_{(j,[x,y])}|]\cdot\log \left(\frac{\mathbb{E}[|\cE_{(j,[x,y])}|]}{\mathbb{E}[|\cE_{(j,[x,y])}\setminus\cK_{(j,[x,y])}|]}\right)+\log^2\log B\right).
	\]
                   			   		
\subsection[Encoding the remaining unknown intervals]{Encoding $\cE\setminus\cK$}
	To encode $\cE$ given $\cK$, we enumerate the good blocks in lexicographical order and write down the $\cE_{(j,[x,y])}(\cK_{(j,[x,y])})$ one by one.
	Let $\cG$ denote the set of good blocks.
	We calculate the expected cost to complete the proof:
	\begin{align*}
		\mathbb{E}\left[\sum_{(j,[x,y])\in\cG}\left(\mathbb{E}[|\cE_{(j,[x,y])}\setminus\cK_{(j,[x,y])}|]\cdot\log \left(\frac{\mathbb{E}[|\cE_{(j,[x,y])}|]}{\mathbb{E}[|\cE_{(j,[x,y])}\setminus\cK_{(j,[x,y])}|]}\right)+\log^2\log B\right)\right].
	\end{align*}
	By applying the law of total expectation, we can restrict the distribution to the event that $\cG\ne\emptyset$.
	We let $(j,[x,y])\sim\cG$ denote that we sample $(j,[x,y])$ from $\cG$ uniformly at random.
	\begin{align*}
		&\mathbb{E}\left[|\cG|\cdot\left(\mathbb{E}_{(j,[x,y])\sim\cG}\left[\mathbb{E}[|\cE_{(j,[x,y])}\setminus\cK_{(j,[x,y])}|]\cdot\log \left(\frac{\mathbb{E}[|\cE_{(j,[x,y])}|]}{\mathbb{E}[|\cE_{(j,[x,y])}\setminus\cK_{(j,[x,y])}|]}\right)\right]+\log^2\log B\right)\right]
		\intertext{Observe that $\mathbb{E}_{(j,[x,y])\sim\cG}\left[\mathbb{E}[|\cE_{(j,[x,y])}\setminus\cK_{(j,[x,y])}|]\right]=\mathbb{E}[|\cE\setminus\cK|]/|\cG|$ and $\mathbb{E}_{(j,[x,y])\sim\cG}\left[\mathbb{E}[|\cE_{(j,[x,y])}|]\right]=\mathbb{E}[|\cE|]/|\cG|$. By \Cref{fact: inequality},}
		\le&\mathbb{E}[|\cE\setminus\cK|]\log(\mathbb{E}[|\cE|]/\mathbb{E}[|\cE\setminus\cK|])+\mathbb{E}[|\cG|\cdot\log^2\log B].
	\end{align*}
	The lemma then follows from the fact that $|\cG|\le p$.

\end{document}